\documentclass[twocolumn, 12pt]{article}

\usepackage[T1]{fontenc} 
\usepackage[english]{babel} 
\usepackage{gensymb} 
\usepackage[a4paper, total={7.2in, 9.72in}]{geometry} 
\usepackage{graphicx} 
\usepackage[]{units}
\usepackage{amsmath, amssymb, amsfonts, amsthm, fouriernc, mathtools}
\usepackage{float}
\usepackage[font=footnotesize]{caption}

\usepackage[switch]{lineno}  

\usepackage{titlesec} 
\titleformat*{\section}{\normalsize \bfseries}
\titleformat*{\subsection}{\normalsize \bfseries}

\titleformat{\subsection}[runin] 
  {\normalfont\normalfont\bfseries}{\thesubsection}{1em}{}

\makeatletter 
\renewcommand\@biblabel[1]{#1} 
\makeatother

\usepackage{titling} 

\pretitle{\Large\bfseries} 
\posttitle{} 
\title{Climate change impacts on large-scale electricity system design decisions for the 21st Century} 
\author{\normalsize S. Kozarcanin$^{1}$, H. Liu $^{1}$ and G. B. Andresen$^{1}$ \\
 \normalsize $^{1}$Department of Engineering, Aarhus University, Denmark.}
\date{\vspace{-5ex}}

\begin{document}

\twocolumn[
  \maketitle]

\newpage

\twocolumn[
  \begin{@twocolumnfalse}
  \section*{Abstract}
      \noindent \textbf{Efforts to reduce climate change, but also falling prices and significant technology developments currently drive an increased weather-dependent electricity production from renewable electricity sources. In light of the changing climate, it is highly relevant to investigate the extent of weather changes that directly impacts the best system design decisions for these weather-dependent technologies. Here, we use three IPCC representative CO$_2$ concentrations pathways for the period 2006--2100 with six high-resolution climate experiments for the European domain. Climate elements are used to calculate bias adjusted 3-hourly time series of wind and solar generation, and temperature corrected demand time series for 30 European countries using state-of-the-art methodology. Weather-driven electricity system analysis methodology is then applied to compare four key metrics of highly renewable electricity systems. We find that climate change does not have a discernible impact on the key metrics of the combined electricity system dynamics, and conclude that the effect on important system design parameters can likely be ignored. } \\
  \end{@twocolumnfalse}
]

\twocolumn[
\section*{Significance statement}
\begin{@twocolumnfalse}
Globally, electricity production from wind and solar sources are increasing significantly. The increase is primarily driven by lower costs, and by political efforts to limit climate change. Climate change, however, may significantly change the weather that drives these sources of renewable energy. We find that the impact of climate change on a future highly renewable European electricity system is small compared to current differences from one weather year to another and compared to differences between designs, e.g., with different levels of international power transmission lines or different mix of wind and solar generators. As a consequence, sound electricity system design decisions and academic studies based on historical weather data will not be strongly affected by climate change.
\end{@twocolumnfalse}
] 

\clearpage
\section*{Introduction}

\begin{figure}[H]
	\centering	
	\includegraphics{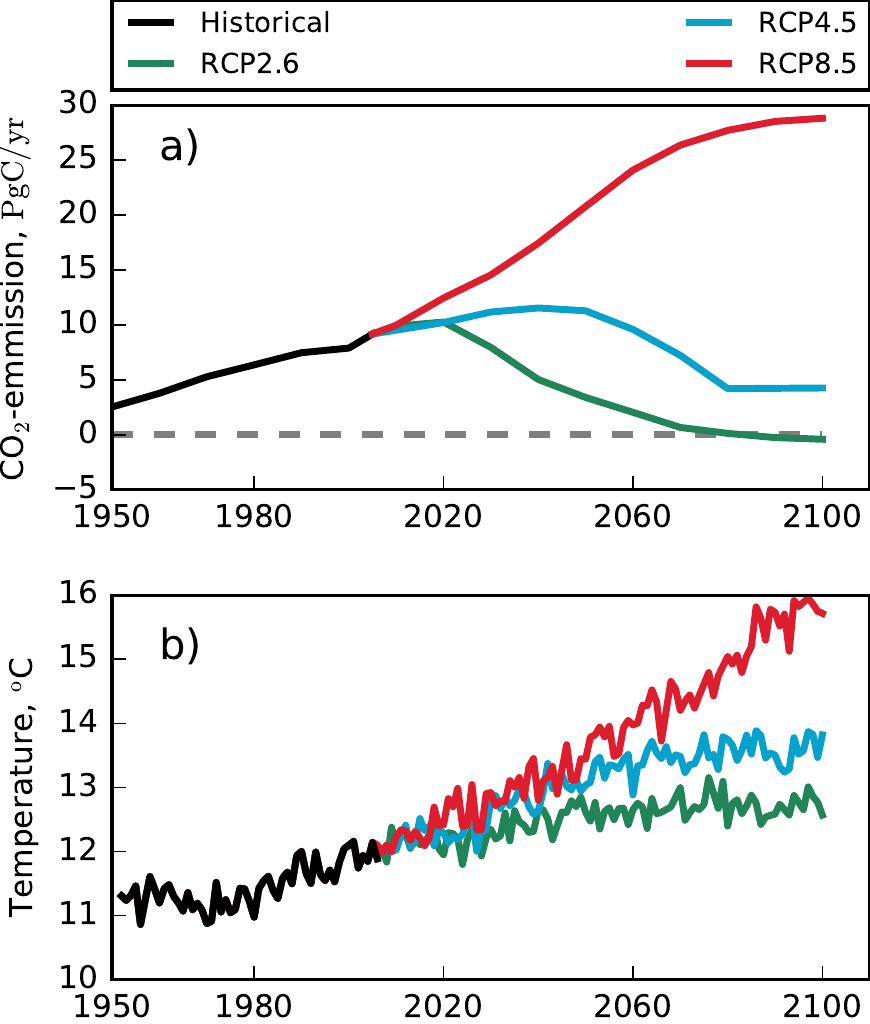}	
	\caption{\textbf{RCP key metrics.} Panel a: CO$_2$ emissions as a function of model year for the various projections of climate outcomes. Panel b: Average near surface temperature over Europe for the regional climate model HIRHAM5-EC-EARTH as a function of model year for identical pathways. Black, green, blue and red projections represent the historical, RCP2.6, RCP4.5 and RCP8.5 pathways.}
	\label{figure:RCPs} 
\end{figure}

\noindent The global climate is currently undergoing vast changes, primarily due to anthropogenic emissions of heat-trapping gases into the atmosphere \cite{myhre2013anthropogenic}. As a consequence, the global average surface temperature has increased since the pre-industrial ages by approximately 0.2~$\degree$C per decade. A similar rate of temperature increase is projected to occur during the first half of the 21st century \cite{hansen2006global}, and at present, the changing climate is showing global impacts on natural and human systems. In this work, we focus on the impacts of a changing climate on the dynamics of a future, highly renewable, large-scale electricity system for Europe. \\


\noindent Research on the impacts of climate change on large scale electricity systems is in its infancy and few studies have been performed in this field \cite{ciscar2014integrated}. For an aggregated European energy system, P. Downling \cite{dowling2013impact} finds outweighing demand side impacts to the supply side on the basis of the POLES energy model. Due to global heating, the cooling demand increases while the heating demand decreases \cite{dowling2013impact}. More specifically, a study conducted by Pilli-Sihvola~\textit{et al.} \cite{pilli2010climate} shows a decreasing need for heating in Central and Northern Europe based on the SRES A2, A1B and B1 emission scenarios. Southern Europe experiences a large increase in cooling demand that in turn overcomes the decrease in heating demand \cite{pilli2010climate}. This finding is in correspondence with a study on combined Germany and Austria for the last quarter of the 21st century \cite{totschnig2017climate}. The spatial distribution of the heating and cooling needs are further reflected in increasing needs for electricity in Southern Europe and decreasing needs in Northern Europe \cite{wenz2017north}. Individual country studies on Norway \cite{seljom2011modelling}, Finland \cite{jylha2015energy}, Slovenia \cite{dolinar2010predicted}, Austria \cite{berger2014impacts} and Switzerland \cite{christenson2006climate} agree upon the latter findings. Spinoni~\textit{et al.} \cite{spinonichanges} shows that if population weighting is applied to the temperature profiles, the degree-day trends do not show remarkable differences during the 21st century for the RCP4.5 and RCP8.5 emission scenarios. \\

\noindent By using the SRES A1B emission scenario, Tobin~\textit{et al.} \cite{tobin2015assessing} and Barstad~\textit{et al.} \cite{barstad2012present} find negligible changes in the wind power potential over the Baltic Sea and the surroundings. Similar results are found by Tobin~\textit{et al.} \cite{tobin2016climate} for Europe by using the RCP4.5 and RCP8.5 emissions scenarios. McColl~\textit{et al.} \cite{mccoll2012assessing} arrive at similar conclusions for the UK, Seljom~\textit{et al.} \cite{seljom2011modelling} for a study-case on Norway based on 10 climate experiments and Koch~\textit{et al.} \cite{koch2015impact} for Germany by using the RCP2.6 and RCP8.5 emission scenarios for the period 2031--2060. Moreover, Schlott~\textit{et al.} \cite{schlott2017} discover increasing wind correlation lengths along with increasing wind power generation in the same regions of Europe by using the RCP8.5 emission scenario. Tobin~\textit{et al.} \cite{tobin2015assessing} and Bloom~\textit{et al.} \cite{bloom2008climate} project a decreasing wind power potential over the Mediterranean area by using the SRES A1B and A2 emission scenarios. Similar findings along with decreasing correlation lengths are shown by Schlott~\textit{et al.} \cite{schlott2017}. For an aggregated EU27, Dowling \cite{dowling2013impact} found a slight increase in the wind power production until mid-century based on the SRES A1B and E1 emission scenarios. For a wind dominated European energy system under the impact of the RCP8.5 emission scenario, increasing needs for conventional power and storage are found in most of the central, north and north-west areas \cite{weber2017impact} with a more homogeneous wind power generation \cite{wohland2017more} at the end of the 21. century. Jerez~\textit{et al.} \cite{jerez2015impact} project a decreasing solar power generation in the Scandinavian area and slight increases over Southern Europe by using the RCP4.5 and RCP8.5 emission scenarios. This is in correspondence with increased means of solar irradiation along with increased solar correlation length in large parts of Southern Europe \cite{schlott2017}. The opposite behaviour is observed for the Northern parts of Europe \cite{schlott2017}. For an aggregated EU27, an increasing amount of solar PV generation is found \cite{dowling2013impact}. Based on 44 electricity scenarios for Europe, Berrill~\textit{et al.} \cite{berrill2016environmental} find that large penetrations of wind and solar power emit the least $CO_2$ within the ensemble of scenarios and as a consequence contribute the most to climate change mitigations. \\ 

\begin{figure*}[h!]
	\centering	
	\includegraphics{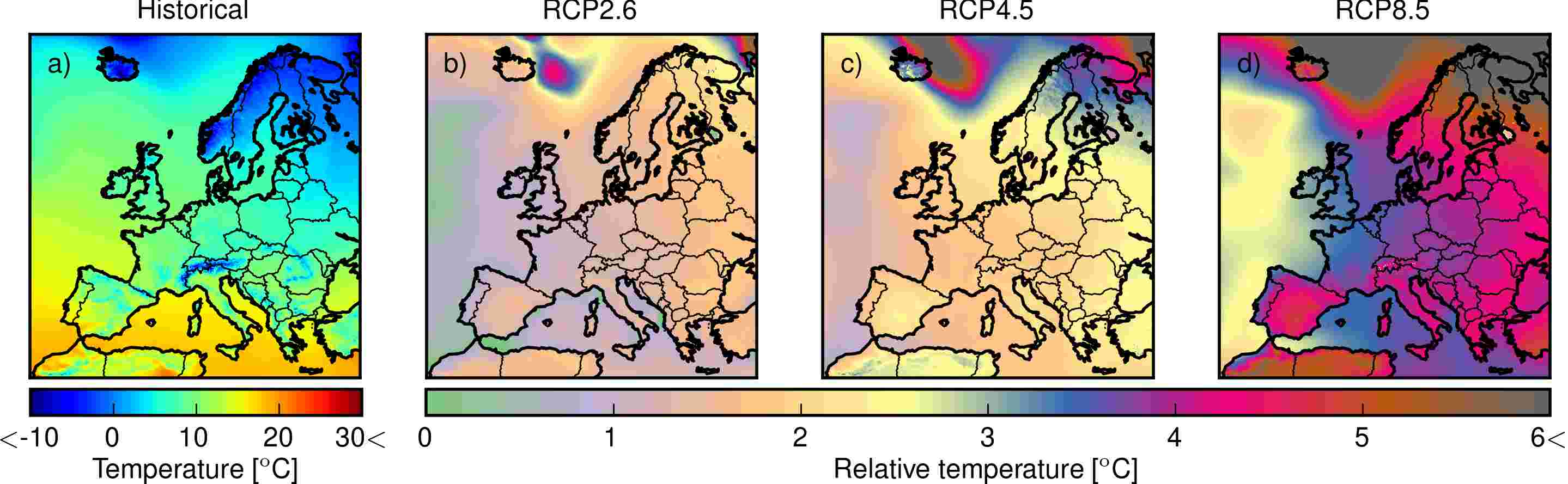}
	\caption{\textbf{20-year average values of the near surface temperature.} The $\unit[2]{m}$ surface temperature is based on the latest generation of IPCC's climate projections, RCP's, for the climate model HIRHAM5-EC-EARTH. Panel a shows the average absolute temperature for the historical period along with a unique colorbar. Panels b,c and d show the temperature increase relative to the historical period and share one colorbar.}
	\label{figure:Maps} 
\end{figure*}

\noindent We add to the literature by investigating the key infrastructure metrics of a fully connected, highly renewable European electricity system by using the latest generation of climate projections provided by the Intergovernal Panel on Climate Change, IPCC \cite{moss2010next}. The climate data used in this study originates from six combinations of four RCMs (Regional Climate Models), downscaling three CMIP (Coupled Model Inter-comparison Project) Phase 5 \cite{taylor2012overview} GCMs (Global Climate Models) under the IPCC representative concentration pathway scenarios, RCPs, see supplementary material. In order to represent a broad range of climate outcomes, three scenarios, RCP2.6 \cite{vuuren2011rcp2}, RCP4.5 \cite{thomson2011rcp4} and RCP8.5 \cite{riahi2011rcp}, have been implemented into the global climate models. \\

\noindent The climate scenarios are heavily influenced by the policy of the 2015 Paris agreement, high $CO_2$ emission cost policies, and no climate policies, respectively. In particular, if the goals of the Paris agreement are to become real, there is an urgent need for short and long-term action in reducing the $CO_2$ emissions, as explained in detail in the Emission Gap Report 2017 \cite{emissiongap2017}. The underlying assumptions of the pathways result in $CO_2$ emissions as shown in Figure~\ref{figure:RCPs}a. RCP2.6 and RCP4.5 project a decrease in the $CO_2$ emissions during the 21st century, while RCP8.5 projects an increase which stagnates at the end of the century. These projections are reflected in the European average temperature for, e.g., the regional climate model HIRHAM5 \cite{christensen2007hirham}, downscaling the global climate model ICHEC-EC-EARTH \cite{hazeleger2012ec}, from now on denoted as HIRHAM5-EC-EARTH, shown in Figure~\ref{figure:RCPs}b. Based on the emission scenarios, 3-hourly data on climate elements with a $0.11\degree$ spatial resolution has been fostered on behalf of the EURO-CORDEX project \cite{kotlarski2014regional, prein2016precipitation}. Figures~\ref{figure:Maps} b, c and d exemplify the 2080--2100 average near-surface temperature over Europe for the future emissions scenarios based on the climate model HIRHAM5-EC-EARTH. Panel~a represents the absolute temperature of the historical period 1986--2006. The most extreme temperature increases reach values of more than $\unit[6]{\degree C}$. \\

\noindent In the present study, the main source of error originates from the performance of the GCMs that provides the boundary conditions for the RCMs used to calculate  the time series of wind and solar power generation as well as electricity demand time series from which the final results are derived. Various studies have engaged in tracking the evolution of CMIP5 GCMs and their predecessor CMIP3 GCMs by comparison to observed data, satellite data or reanalysis. In the following, we summarize the quality of the climate elements that are most important for the present study. 

\noindent Iqbal~\textit{et al.} \cite{iqbal2017analysis} investigates the North Atlantic eddy-driven jet stream in 11 CMIP5 GCMs by evaluating the jet latitudes and wind speeds in a historical run from 1980-2004. Compared to reanalysis, typical biases between $\unit[1]{\degree}$  and $\unit[2]{\degree}$ are observed in the latitude mean seasonal cycle anomaly for a majority of the GCMs. The corresponding seasonal amplitude variations reach up to $\pm \unit[10]{\degree}$ for a few GCMs, which is an over or underestimation of about $\unit[5]{\degree}$ compared to reanalysis. But compared to identical CMIP3 GCMs, it is an improvement \cite{hannachi2013behaviour}. In addition, Iqbal~\textit{et al.} \cite{iqbal2017analysis} shows a slight poleward shift in the jet stream under the IPCC RCP4.5 and RCP8.5 induced global warming during the end of the century. This is further confirmed for 26 CMIP5 GCMs under the IPCC RCP8.5 emission scenario, which shows a poleward jet shift of approximately $\unit[2]{\degree}$ in the Southern Hemisphere and $\unit[1]{\degree}$ in the North Atlantic and North Pacific \cite{barnes2013response}. In all GCMs, the mean seasonal anomaly bias for the jet stream wind speeds is overestimated for the winter months while the opposite is the case for summer months. A majority of the GCMs are overestimating the corresponding amplitudes \cite{iqbal2017analysis}. The jet wind speed biases are smaller in the CMIP3 GCMs compared to the CMIP5 GCMs \cite{hannachi2013behaviour}. \\

\noindent The GCM cloud simulations, which most directly affect the solar energy yield calculated in our study, remain the major source of inaccuracy in climate predictions. Lauer~\textit{et al.} \cite{lauer2013simulating} evaluates the total amount of cloud cover, amongst other climate elements, in 27 CMIP5 GCMs by comparing to satellite data covering the years 1986-2007. It is shown that the linear correlation coefficient of the GCM wise 20 year annual average of the total cloud amount ranges from 0.11 to 0.83 with a root mean square error of 10\% to 23\%. For CMIP5 experiments, the IPCCs Fifth Assessment Report AR5 \cite{collins2013long} presents a difference of $\unit[2]{\%}$ to $\unit[3]{\%}$ in regional changes of cloud fractions for the years 2081--2100 under the RCP4.5 emission scenario. For the RCP8.5 emissions scenario, a difference of $\unit[5]{\%}$ to $\unit[6]{\%}$ is presented. \\

\noindent The global radiation budget biases that are most strongly influenced by clouds are the incomming radiative shortwave flux at the surface along with the outgoing flux at top of the atmosphere and the reflected flux. For a number of CMIP5 GCMs, Li \textit{et al.}~\cite{li2013characterizing} finds regional biases ranging from $\unit[-25]{Wm^{-1}}$ to more than $\unit[30]{Wm^{-1}}$ for the incomming flux compared to EBAF-Surface flux radiation product for a time period 2000-2010. The multimodel global area average bias is reduced by $~30\%$ compared to CMIP3 GCMs. Similar findings are evident for the regional outgoing flux as well as reflected flux with an improvement of the multimodel global area average bias with a factor of 2 compared to CMIP3 GCM. With these consideration in mind, it is evident that CMIP5 GCMs have improved since phase 3, although room for further improvement remains. In this work we are compensating for static biases in the solar energy yield as explained in the methods section.\\

\noindent In the present study, surface air temperatures are used to adjust the electrical demand for changes in heating and cooling needs. The GCM ability to accurately replicate this field is receiving increasing attention. A recent study by Cattiaux~\textit{et al.}~\cite{cattiaux2013european} on the European domain shows negatively biased winter temperatures in the North compared to ground observations \cite{van2015international}. Positively biased summer temperatures are observed in the East and Central Europe for 33 CMIP5 GCMs. For the total European domain, the GCM ensemble mean bias is approximately $\unit[-1]{\degree C} \pm \unit[8]{\degree C}$ during winter months. The corresponding values for summer months are $\unit[0.5]{\degree C} \pm \unit[6]{\degree C}$. Similar trends are found for the Northern Eurasia where the winter and summer periods show the largest biases \cite{miao2014assessment}. Small improvements have been made since CMIP3 GCMs \cite{meehl2007global}.\\

\noindent The GCM errors are taken into account by evaluating this study for a set of three different models. In the results section we compare the main results for the ensemble of GCMs. Relatively small deviations are observed which leads to the conclusion that the current GCMs are highly suitable for this study. \\

\section*{Methods}

This section describes the method of the electricity system modelling, data conversion and validation, and the electricity system key metrics. For a detailed descriptions of these quantities one is referred to the supplementary information (SI) 2.1, 2.2 and 2.3\\

\noindent \textbf{Weather-driven electricity system modelling} was introduced for the first time by Heide~\textit{et al.} \cite{heide2010seasonal}. It is a type of top-down analysis that is particularly useful for isolating and understanding the impact of weather dynamics on an electricity system with a large penetration of variable sources, e.g, wind and solar. But unlike most technology-rich bottom-up models, it is not well-suited for determining, e.g., the fuel mix of conventional dispatchable production units. In this paper, we implement a simplified large-scale European electricity system, of which the supply side consists solely of wind and solar power generation as well as a generic dispatchable power source, e.g., dammed hydro power or gas turbines. The demand side consists of national electricity demand. This type of electricity system aggregates the country-wise wind and solar power generation profiles into system-wise profiles. The same applies for the electricity consumption profiles. The effect of power transmission in this system was introduced by Rodriguez~\textit{et al.} \cite{rodriguez2015localized, rodriguez2014transmission}. \\

\noindent The effect of climate change enters the modelling via the weather-dependent wind and solar generation time series as well as the temperature dependent part of the electricity demand. State-of-art methodology has been used to convert raw wind and solar climate data into 3-hourly country-wise wind and solar capacity factor time series. These have been validated for the historical period 1986--2006 against already bias adjusted capacity factors provided by Renewables.ninja \cite{staffell2016using, pfenninger2016long} by minimizing the relative entropy (see SI 3.1 and 3.2).  Hourly country-wise electricity consumption profiles have been provided by the European Network of Transmission System Operators for Electricity, ENTSO-e \cite{entsoe}. These have been corrected for effects of heating and cooling by using the degree day method (see SI 4). Additional detailed descriptions of these methods are found in the supplementary material.  \\

\noindent In the work presented here, wind and solar generator capacities have been scaled such that the EU-wide average generation over the historical period matches the average demand for electricity. This choice has previously been shown to describe the dynamics of a highly renewable electricity system well \cite{heide2010seasonal}. Due to the state of fluctuating weather events and electricity consumption behaviour, the generation and load profiles differ in most time steps. This difference is referred to as the generation-load mismatch, which is defined formally in the methods section. For high renewable penetrations, the generation-load mismatch means that surplus of the variable renewable electricity, VRE, will sometimes be available. This surplus may either be transported to other locations, used by new flexible consumption, stored for later use or simply curtailed. At other times, VRE generation is lacking and an ancillary system must balance the difference between supply and demand, e.g., by a combination of conventional power plants, dispatchable renewable electricity such as dammed hydropower, and electricity storage systems. Variations in instantaneous generation-load mismatch across the continent drive the transmission of renewable surplus to regions with a deficit. Furthermore, the maximum negative values of the mismatch describe the need for dispatchable power capacity, and variations in the mismatch describe the need for on-demand flexible reserves.\\

\noindent \textbf{Key Metrics}. The above considerations give rise to four key metrics that describe a highly renewable electricity system. These are all described formally in the methods section. The first key metric is the average \emph{dispatchable electricity} delivered by the dispatchable power generators for a fixed wind and solar penetration. It is a good indicator of the utility value of the VRES, i.e. the amount of VRE that can directly cover the demand. This metric was first introduced by Heide~\textit{et al.} \cite{heide2011reduced}.\\

\begin{figure}
	\centering	
	\includegraphics{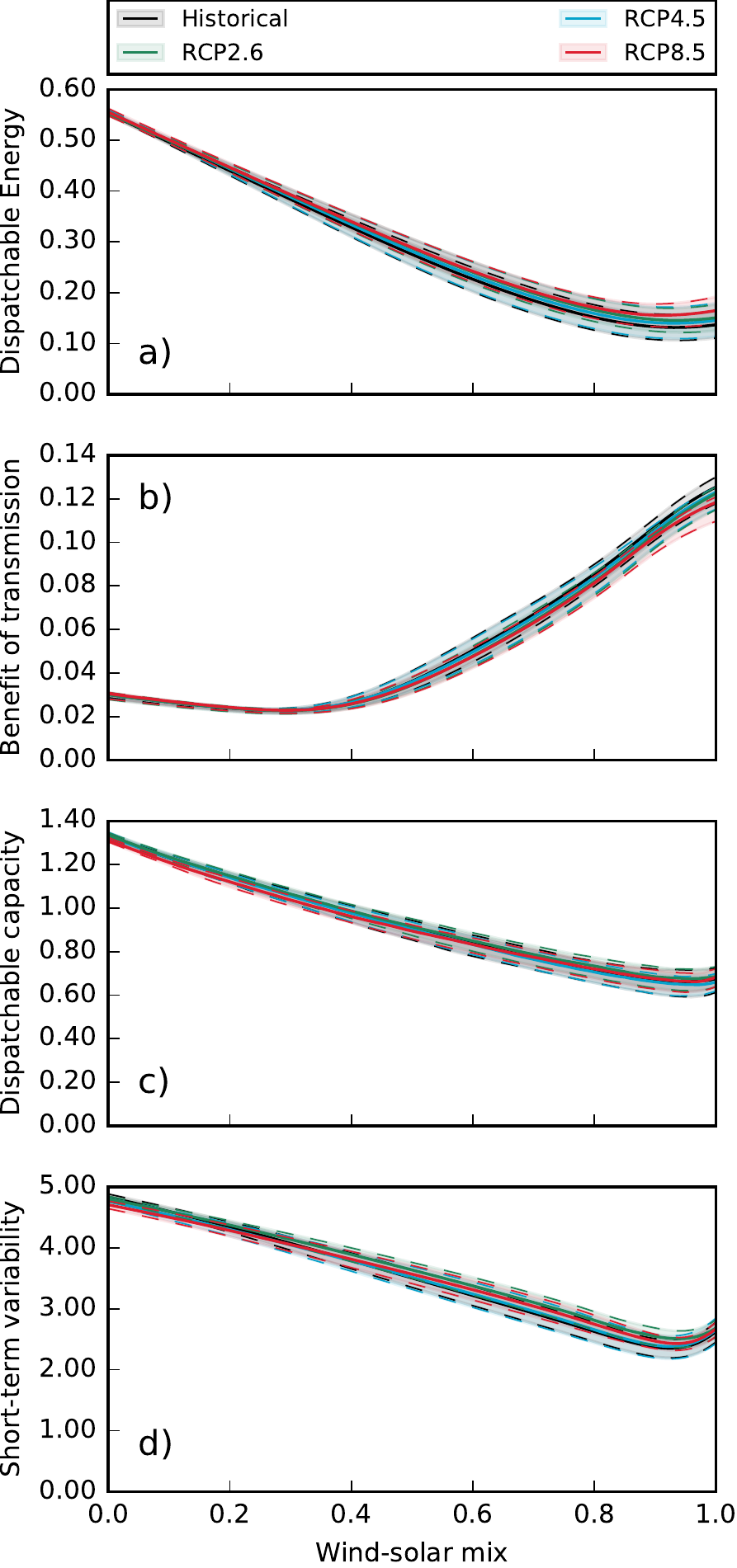}
	
	\caption{\textbf{Key metrics as a function of the wind-solar mix for the historical and end-of-century periods.} The historical period covers 1996--2006 (black) and end of the century period covers 2080--2100 for the RCP2.6 (green), RCP4.5 (blue) and RCP8.5 (red) pathways. A wind-solar mix equal to zero represents a solar only electricity system and wind-solar mix of 1 represents a wind only electricity system. The averages of the annual values are indicated with fully drawn lines, and the corresponding ranges (one sigma) are shown with dashed lines. All key metrics are unitless as described in the methods section.}
	
	\label{figure:KeyMetricsVsAlpha} 
\end{figure}

\noindent Generally, the useful electricity can be increased by means of geographical dispersion, e.g., by combining national systems to smooth out variations in both demand \cite{corcoran2012effects} and renewable generation \cite{becker2014features}. We measure this effect with the second metric which is defined as the absolute difference between balancing electricity required with and without unlimited transmission in a European power grid. We call the metric, the \emph{absolute benefit of transmission} in contrast to the relative benefit of transmission used by Becker~\textit{et al.} \cite{becker2014transmission}. In this paper we do not detail the required transmission capacities, but note that many other studies find that it is cost efficient to build enough to harvest most of this benefit \cite{becker2014transmission, schaber2012transmission}.\\

\noindent The two first measures are concerned with the need for dispatchable electricity, which is closely related to, e.g., CO2 emissions. The third and fourth key metrics indicate what is required to stabilize an electricity system based on VRES and maintain security of supply. The third metric is measured by the maximum \textit{dispatchable capacity} required during low VRE and high demand periods. A number of studies show that VRES only provides very limited firm capacity as in Rodriguez \textit{et. al.} \cite{rodriguez2015localized}. However, this is not a good key metric as it does not capture the correlation between VRES and demand. The fourth and final key metric is the 3-hourly \emph{short-term variability} of the balancing time series. This measure was shown to be a good proxy for the need for on-demand reserve capacity \cite{holttinen2010variability}. The variability is largely caused by meso-scale turbulence patterns that have been shown to describe the spatio-temporal characteristics of wind and solar power generation well \cite{olauson2016restoring}.\\

\noindent We note that the effect of many types of extreme weather conditions are implicitly captured by the key metrics that are calculated from the 3-hourly generation-load mismatch. For instance, an extended period of very hot weather in the RCP8.5 scenario would increase the use of dispatchable electricity and capacity as electricity demand for cooling would increase and the performance of solar panels would decrease compared to the historical scenario. Likewise, strong storms would cause wind turbines in areas with wind speeds exceeding, typically, 25~m/s to cut out abruptly, causing an increased short-term variability as well as increased needs for dispatchable electricity and capacity.\\


\section*{Results}

Initially, we present results only based only on the climate model HIRHAM5-EC-EARTH. Towards the end, we present the robustness of the results by comparing similar findings from all six climate models. We stress, however, that in the analysis, all six climate models have been treated identically.\\

\noindent Overall, the different climate scenarios show a minor impact on the key metrics of a highly renewable European electricity system. This is illustrated in Figure~\ref{figure:KeyMetricsVsAlpha}, where the key metrics are shown as a function of the wind-solar mix for the end of the century period 2080--2100, where all RCPs are most developed. We observe that the difference between average values for different climate scenarios is smaller than the corresponding annual variation (one sigma) in nearly all cases. In other words, annual variations within a given climate change scenario are larger than the difference between scenarios. A paired t-test reveals that, in most cases, the distributions of annual scores cannot be assumed to come from different underlying distributions, indicating that the null-hypothesis that climate change has no impact on the key metrics cannot be rejected (95\% confidence). To some extend this finding is related to the choice of generator capacities relative to demand because a decrease in demand is balanced by an independent decrease in consumption. This is evident from Table~\ref{table:RCPimpact}, where it can be seen that both annual demand and wind and solar generation are decreasing slightly in the future scenarios. At a 95\% significant level, the annual demand and annual solar capacity factors can both be considered different for the different climate change scenarios. The wind capacity factors, on the other hand, cannot be considered different. All test scores are listed in the supporting material.\\

\begin{table*}[h!]
\centering
\begin{tabular}{cccccc}
\hline
& Historical & RCP2.6 & RCP4.5 & RCP8.5 \\
$\langle W \rangle _{\text 20yr}$ & $1.00 \pm 2.56\cdot 10^{-2}$ & $0.98 \pm 2.39\cdot 10^{-2}$ & $0.99 \pm 3.65\cdot 10^{-2}$ & $0.96 \pm 2.62\cdot 10^{-2}$  \\
$\langle S \rangle _{\text 20yr}$ & $1.00 \pm 1.36\cdot 10^{-2}$ & $0.99 \pm 2.14\cdot 10^{-2}$ & $0.98 \pm 1.85\cdot 10^{-2}$ & $0.97 \pm 3.00\cdot 10^{-2}$  \\
$\langle L \rangle _{\text 20yr}$ & $1.00 \pm 4.59\cdot 10^{-3} $ & $1.00 \pm 3.40\cdot 10^{-3}$ & $0.99 \pm 4.68\cdot 10^{-3}$ & $0.98 \pm 2.70\cdot 10^{-3}$   \\
\hline 
\end{tabular}
\caption{\textbf{20--year average values of the normalized wind and solar power along with the normalized consumption.} }
\label{table:RCPimpact}

\end{table*}

\noindent The difference between the climate change scenarios can also be compared to the difference between systems with different wind-solar mix within a scenario. This comparison reveals to what extend it is more important to choose the right wind-solar mix independently of RCPs, or if it is most important to design the electricity system for the right RCP. For the dispatchable electricity key metric shown in Figure~\ref{figure:KeyMetricsVsAlpha}a, we find the largest difference between climate scenarios for a wind-solar mix of 1.0. The RCP8.5 scenario has an average value of 0.165 while the corresponding number for the historical period is 0.137. However, within each of the two scenarios, the difference between a wind dominated and a solar dominated mix is significantly larger. This means that in a planning context it is more important to consider the wind-solar mix than the degree of climate change. Similar arguments can be made for the other key metrics, i.e. the benefit of transmission, the dispatchable capacity and the short-term variability.\\

\noindent A solar-dominated electricity system requires dispatchable electricity that is slightly higher than half of the electric consumption. This is explained by the day/night pattern in which little dispatchable electricity is needed during day time and full dispatchable electricity needs are required for the night time. On the other hand, the production of wind power during both day and night time reduces the need for dispatchable electricity to less than 20\% of the annual demand. These findings are in correspondence with earlier work \cite{heide2011reduced}. The benefit of transmission is highly dependent on the spatial correlation lengths of the wind speeds and solar irradiance \cite{widen2015variability}. The high correlation lengths of the solar irradiance due to the systematic variability of the sun's position in the sky give rise to low transmission benefits. Due to the production of PV power over large areas, minor needs for transmission are present. On the other hand, the smaller wind speed correlation length of approximately $\unit[600-1000]{km}$, leads to areas with high wind electricity production and areas with low wind power production. Under these circumstances, the need for transmission increases. The short-term variability for solar domination shows higher values in contrast to wind domination. This behaviour is due to stronger intra-day temporal variations in the solar irradiance compared to temporal variations in the wind speeds. It is further reflected in the dispatchable capacity. \\

\begin{figure*}
	\centering	
	\includegraphics{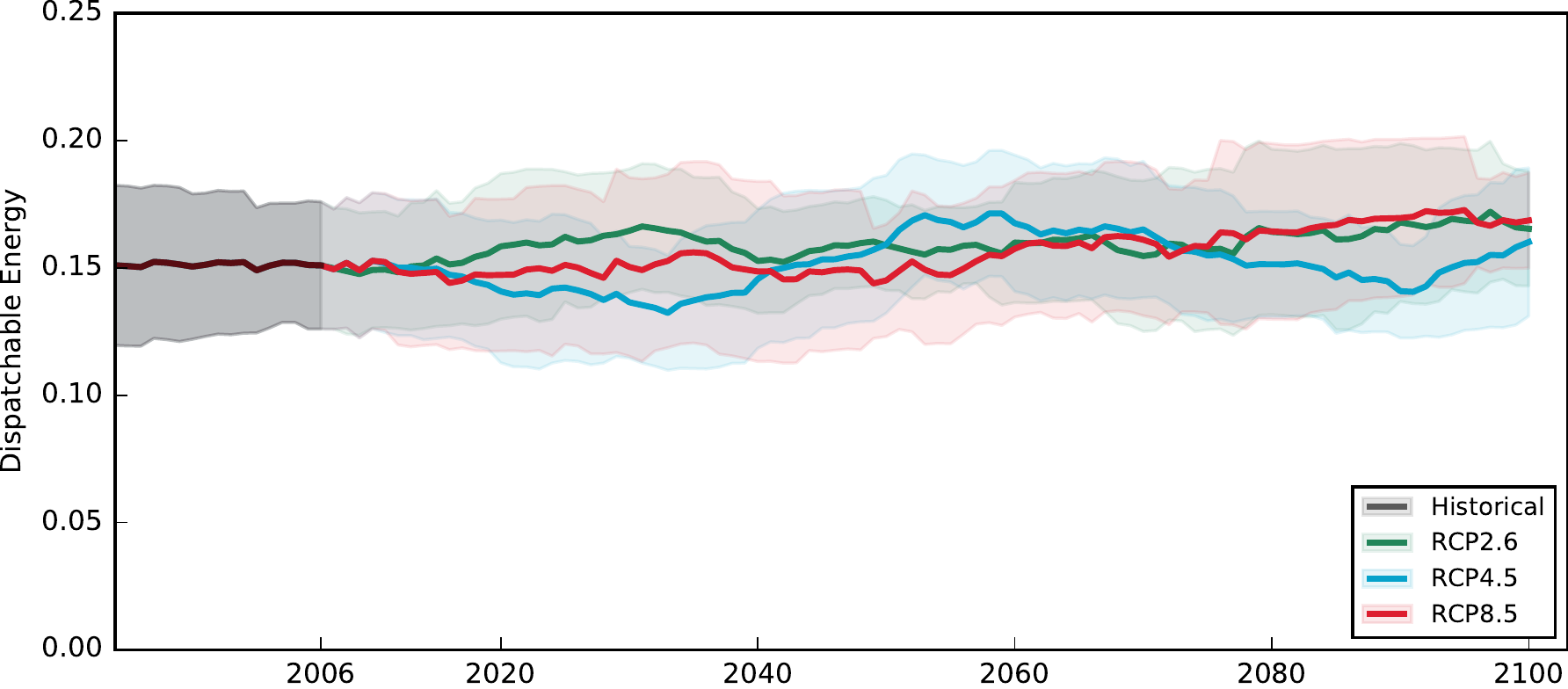}
	\caption{\textbf{Evolution of the dispatchable electricity as a function of model year for the climate model HIRHAM5-EC-EARTH.} The historical period covers 1996--2006 (black) and the future scenarios RCP2.6 (green), RCP4.5 (blue) and RCP8.5 (red) covers the years 2006--2100. The 20 year averages of the annual values are indicated with fully drawn lines, and the corresponding ranges of the one sigma standard deviation are shown with shaded areas. All key metrics are unitless as described in the methods section.}
	\label{figure:KeyMetricsVsTime} 
\end{figure*}

\noindent A number of studies focus on the end of the century period alone. However, we also explore the evolution of the key metrics towards the end of the century. In the following we focus on the dispatchable electricity for a wind-solar mix of 0.8, shown in Figure~\ref{figure:KeyMetricsVsTime}. Similar results can be found for the other key metrics in the supporting material. For this case, we do not find a gradual transition from the historical reference period towards the end of the century. Instead, the trajectories of the RCPs change over time such that, e.g., the 20-year average values of each RCP cross the other trajectories multiple times. This means that the general conclusions about the relative importance of climate change on the electricity system, drawn above for the end of the century period, also hold for the transition, and it is not possible to regard specific numerical differences between RCPs as the product of a gradual evolution towards a stable climatic end-point. Rather the 20--year averages are not stable to an extent where they can be regarded as different between RCPs. A longer time window could be used instead. However, since the climate is changing during the course of the century, see Figure~\ref{figure:RCPs}, long periods do not represent a stable climatic situation.\\

\begin{figure*}
	\centering	
	\includegraphics{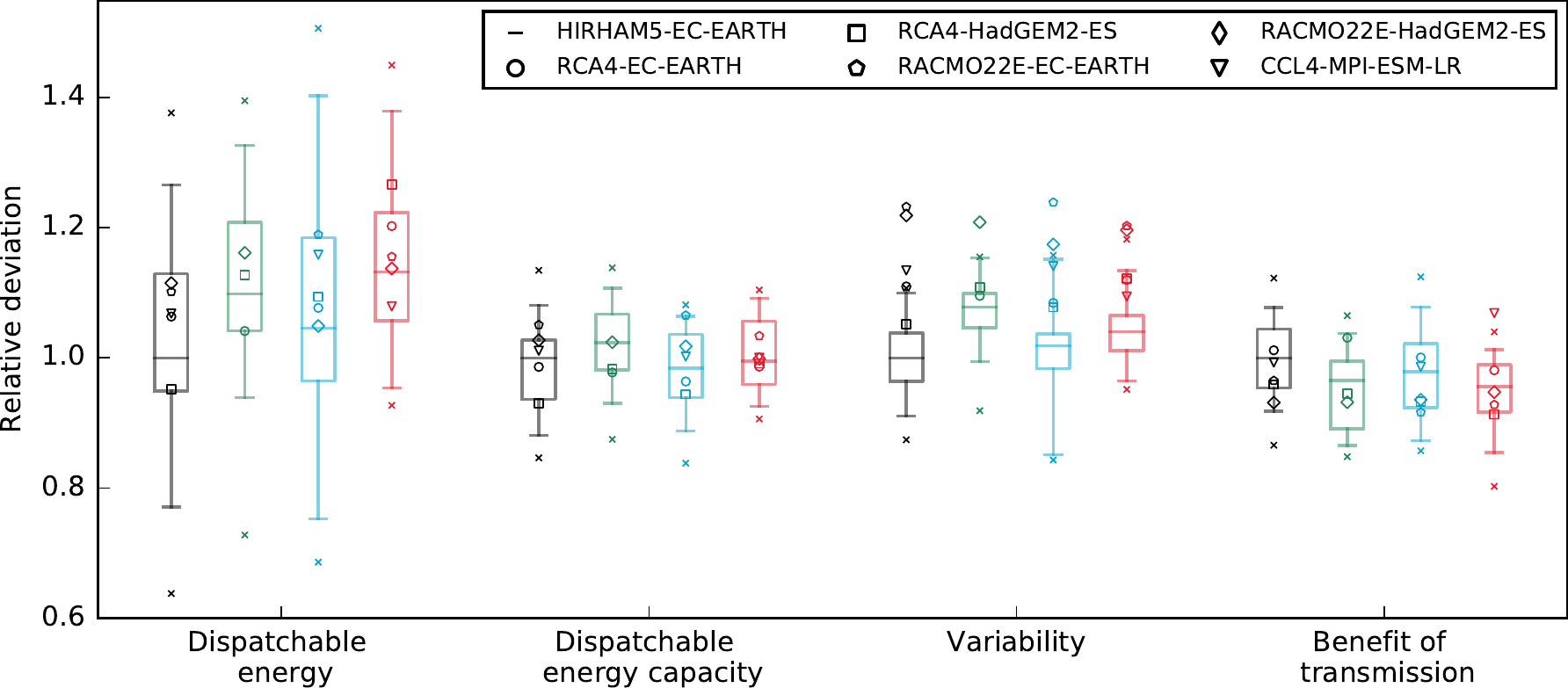}
	
	\caption{\textbf{Key metrics for six regional climate models from the EURO-CORDEX project \cite{kotlarski2014regional, prein2016precipitation}.} These are compared for the historical period 1996--2006 (black) and end of the century 2080--2100 for the RCP2.6 (green), RCP4.5 (blue) and RCP8.5 (red) scenarios. Annual values for the HIRHAM5-EC-EARTH \cite{christensen2007hirham} model are indicated as box plots. For the remaining models, only the median values of the annual means are shown. All key metrics are unitless as described in the methods section.}
	
	\label{figure:ModelComparison} 
\end{figure*}

\noindent Above, we have discussed results based on the HIRHAM5-EC-EARTH climate model. Now we turn to the five other combinations of regional and global climate models included in the study in order to establish to what extent the results are in agreement across the models. In general, we do not find identical numerical results for all models despite the fact that all time series have been bias adjusted on the same historical period against the same reference. However, the overall findings remain, independently of the models. Figure~\ref{figure:ModelComparison} shows a comparison of the four electricity system key metrics for the different models in the historical period as well as in the end of the century period for the RCPs. In most cases, the 20-year average values fall between the first and the third quartile of the annual variation, which means that the difference between individual RCMs is smaller than the annual difference within a given RCM. For the variability, this is not the case. The main reason for this is that the wind and solar time series are bias adjusted such that the distribution of instantaneous values matches the reference. However, the variability depends on the correlation between consecutive hours. This difference is not directly subject to correction, which means that models are likely to behave more differently on this parameter.\\

\section*{Conclusion}

\noindent The different climate change scenarios show a minor impact on the key metrics of a highly renewable European electricity system. On the supply side, both wind and solar power generation are affected in a way that generally reduces performance slightly as their average electricity output decreases and their variability increases with an increasing effect of climate change. However, temperature changes affect the demand for electricity stronger. In most cases, this has an opposing effect as winter demands, primarily in Northern Europe, are reduced due to the increased temperatures. As an example, increasing dispatchable electricity needs in the pathway with the most extreme climatic changes, RCP8.5, due to lower yields from VRES turn into decreasing needs due to lower electricity demands, and could most certainly be either reduced or increased further by impacts from sources not related to climatic conditions such as electricity efficiency measures, population change or increased electrification of, e.g., heating and transportation sectors \cite{jacobson2015100}. The findings above are in agreement with a study on Southern Africa which also shows negligible effects of climate change on the wind and solar industry \cite{fant2016impact}.\\

\noindent In designing a future highly renewable electricity system that is robust against climate change, it is important to focus on reaching a mixture of wind and solar power generation that minimizes the need for dispatchable electricity as the climate, during the next century, does not bring prominent changes to the electricity system infrastructure key metrics. \\

\noindent An important consequence of these findings is that for most purposes, it is not required to take into account the effect of climate change on the VRES wind and solar when designing future highly renewable electricity systems. It is advisable, however, to consider changes in the electricity demand due to climatic changes as these may indirectly affect the best generator mix, e.g., by decreasing the winter demand and increasing the summer demand. Should a stronger sector coupling, in particular between the heating, cooling and electricity sectors, become a reality, these climatic effects would be increased.\\

\section*{Acknowledgement}
The authors wish to express their gratitude to O. B. Christensen and F. Boberg from the Danish Meteorological Institute for their contribution to understanding the climate outcomes and for supplying out data from the regional climate model HIRHAM5. G. Nikulin from the Swedish Meteorological and Hydrological Institute is thanked for supplying out data from the regional climate model RCA4. E. van Meijgaard from the Royal Netherlands Meteorological Institute is thanked for supplying out data from the regional climate model RACMO22E. P. Lenzen from the German Climate Computing Center is thanked for supplying out data from the regional climate model CCLM4. Professor M. Greiner is thanked for fruitful discussions. A sincere thank to Mette Stig Hansen for her diligent proof reading of this paper. Thanks to Heidi S\o ndergaard for giving graphical inputs to the figures. Finally, thanks to the Aarhus University Research Foundation (AUFF) for providing financial support.

\section*{Correspondence} Correspondence and requests for materials should be addressed to G. B. Andresen (email: gba@eng.au.dk) or S. Kozarcanin (email: sko@eng.au.dk)

\bibliography{bib}

\end{document}